\documentclass[Physsubmission, Phys]{SciPost}

\usepackage{float}
\usepackage{floatflt}

\hypersetup{
    colorlinks,
    linkcolor={red!50!black},
    citecolor={blue!50!black},
    urlcolor={blue!80!black}
}

\usepackage[bitstream-charter]{mathdesign}
\newcommand{\gsim}{\raisebox{-0.07cm}{$\, \stackrel{>}{{\scriptstyle
\sim}}\, $}}
\newcommand{\GeV}{\rm GeV}

\DeclareSymbolFont{usualmathcal}{OMS}{cmsy}{m}{n}
\DeclareSymbolFontAlphabet{\mathcal}{usualmathcal}

\begin{document}

\begin{center}{{
DESY 21--092, DO--TH 21/20, TTP 21--023, SAGEX--21--11, RISC Report Series 21--14}\\ \Large 
\textbf{
New 2– and 3–loop heavy flavor corrections to unpolarized and
polarized deep-inelastic scattering\footnote{Also contribution to the Proceedings of RADCOR 2021, Tallahassee, 
FL, May 2021.}
\\
}}\end{center}

\begin{center}
J. Ablinger\textsuperscript{1},
J. Bl\"umlein\textsuperscript{2$\star$},
A. De Freitas\textsuperscript{2},
M. Saragnese\textsuperscript{2}, 
C. Schneider\textsuperscript{1},
K. Sch\"onwald\textsuperscript{3},
\end{center}

\begin{center}

{\bf 1} {
Johannes Kepler University Linz,
Research Institute for Symbolic Computation (RISC),\\
Altenberger Stra\ss{}e 69,
A-4040, Linz, Austria}
\\
{\bf 2} { Deutsches Elektronen-Synchrotron DESY,\\
Platanenallee 6, D-15738 Zeuthen, Germany}
\\
{\bf 3}
{Institut f\"ur Theoretische Teilchenphysik Campus S\"ud,\\
Karlsruher Institut f\"ur Technologie (KIT), D-76128 Karlsruhe, Germany}

\vspace*{3mm}

* Johannes.Bluemlein@desy.de
\end{center}

\begin{center}
\today
\end{center}


\definecolor{palegray}{gray}{0.95}
\begin{center}
\colorbox{palegray}{
  \begin{tabular}{rr}
  \begin{minipage}{0.1\textwidth}
    \includegraphics[width=22mm]{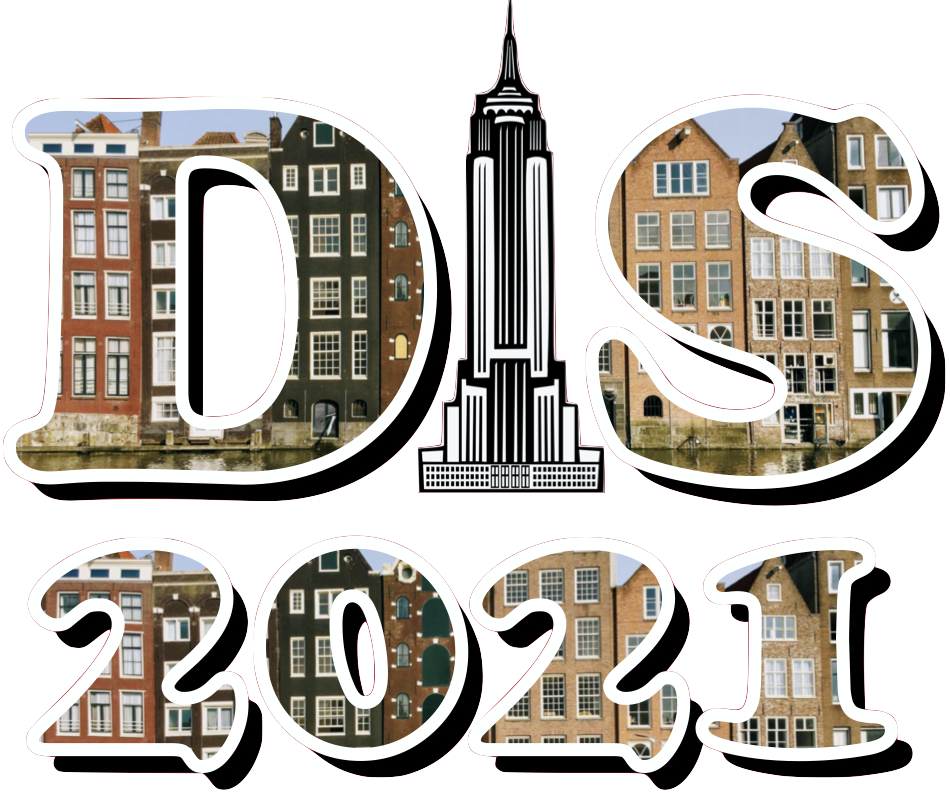}
  \end{minipage}
  &
  \begin{minipage}{0.75\textwidth}
    \begin{center}
    {\it Proceedings for the XXVIII International Workshop\\ on Deep-Inelastic Scattering and
Related Subjects,}\\
    {\it Stony Brook University, New York, USA, 12-16 April 2021} \\
    \doi{10.21468/SciPostPhysProc.?}\\
    \end{center}
  \end{minipage}
\end{tabular}
}
\end{center}

\section*{Abstract}
{\bf
A survey is given on the new 2-- and 3--loop results for the heavy flavor contributions to deep--inelastic 
scattering in the unpolarized and the polarized case. We also discuss related new mathematical aspects
applied in these calculations.
}

\vspace{10pt}
\noindent\rule{\textwidth}{1pt}
\tableofcontents\thispagestyle{fancy}
\noindent\rule{\textwidth}{1pt}
\vspace{10pt}

\section{Introduction}
\label{sec:1}

\vspace*{1mm}\noindent
The scaling violations of deep--inelastic structure functions provide a precise way to measure the 
strong coupling 
constant $\alpha_s(M_Z)$ \cite{Bethke:2011tr}. This requires to calculate their $Q^2$ dependence both due to 
the massless and massive contributions at highest precision. Moreover, the structure functions also allow the 
precise measurement of the charm quark mass, $m_c$, \cite{Alekhin:2012vu}. At future facilities, like the EIC
\cite{Boer:2011fh} or the LHeC \cite{LHEC}, operating at high luminosity, unpolarized and polarized structure 
functions can be measured at high precision to supplement and extend the present deep--inelastic 
world data. The massless 
QCD corrections are available to 3--loop order \cite{Moch:2004pa,Vogt:2004mw,Ablinger:2014nga,Moch:2014sna,
Moch:2015usa,Ablinger:2017tan,Behring:2019tus,Blumlein:2021enk,Vermaseren:2005qc}. In the case of massive 
corrections, analytic results at the 3--loop level can currently only be obtained in the asymptotic approximation,
$Q^2 \gg m_q^2$, with $Q^2$ the virtuality of the process and $m_q$ the heavy quark mass, 
cf.~\cite{Buza:1995ie},
which, however, are accurate to $\sim 1\%$ for the structure function $F_2(x,Q^2)$ in the heavy 
quark region of smaller values of $x$ for $Q^2/m_q^2 \gsim 10$
already. This region should be chosen also to avoid higher twist effects, requiring at least the cuts $W^2 > 
15~\GeV^2, Q^2 > 10~\GeV^2$, \cite{Alekhin:2012ig}.

After having obtained a series of Mellin moments for massive 3--loop operator matrix elements (OMEs) in 2009 
\cite{Bierenbaum:2009mv} the systematic calculation of the different OMEs contributing in the unpolarized and 
polarized case for the massive Wilson coefficients and the OMEs contributing to the matching conditions in the 
variable flavor number scheme (VFNS) has been started for the single mass \cite{Ablinger:2014nga,Ablinger:2010ty,
Ablinger:2014lka,Ablinger:2014vwa,Ablinger:2014uka,GLUE,Behring:2014eya,Blumlein:2014fqa,Behring:2015roa,
Behring:2015zaa,Behring:2016hpa,Ablinger:2017ptf,Ablinger:2019etw,Behring:2021asx,Blumlein:2021xlc}
and the two--mass contributions \cite{Ablinger:2017xml,Ablinger:2018brx,Ablinger:2020snj,Ablinger:2019gpu}.

At 2--loop order the unpolarized and polarized non--singlet and pure singlet contributions have been calculated 
for the full kinematic region and analytic results have been derived for the power corrections in 
Refs.~\cite{Buza:1995ie,Buza:1996xr,Blumlein:2016xcy,Blumlein:2019qze,Blumlein:2019zux} to the structure functions
$F_2, F_L$ and $g_1$.

In Section~\ref{sec:2} we describe the status of the calculation of the massive 3--loop OMEs and discuss the 
calculation methods used in Section~\ref{sec:3}.  Recent results in the single mass and two--mass cases are
reported in Sections~\ref{sec:4} and \ref{sec:5} and Section~\ref{sec:6} contains the conclusions. 
\section{Status of the massive OME calculations}
\label{sec:2}

\vspace*{1mm}\noindent
In the leading twist approximation deep--inelastic structure functions have the representation
\begin{eqnarray}
F_{2,L}(x,Q^2) = \sum_{i,q} \mathbb{C}_{2,L}^{(i)} \left(x, \frac{Q^2}{\mu^2}, 
\frac{m_q^2}{\mu^2}\right) \otimes 
f_{(i)}(x, \mu^2),
\end{eqnarray}
with $\otimes$ the Mellin convolution. A corresponding relation holds for the polarized structure function
$g_1(x,Q^2)$.
Here $\mathbb{C}^{(i)}$ denotes the Wilson coefficient related to the parton density $f_{(i)}$, with $\mu$ the 
factorization scale and $m_q$ the heavy quark masses $m_q = m_c, m_b$. $x$ denotes the Bjorken 
variable.
The Wilson coefficients can be decomposed 
into the massless, $C$, and massive contributions, $H$,
\begin{eqnarray}
\mathbb{C}_{2,L}^{(i)} \left(x, \frac{Q^2}{\mu^2}, \frac{m_q^2}{\mu^2}\right) =
{C}_{2,L}^{(i)} \left(x, \frac{Q^2}{\mu^2}\right) + 
H_{2,L}^{(i)}\left(x, \frac{Q^2}{\mu^2}, 
\frac{m_q^2}{\mu^2}\right).
\end{eqnarray}
At large scales $Q^2$ the heavy flavor Wilson coefficients have the representation 
\cite{Buza:1995ie}
\begin{eqnarray}
H_{2,L}^{(i)}\left(x, \frac{Q^2}{\mu^2}, \frac{m_q^2}{\mu^2}\right) =
\sum_{j,q} C_{(j),2,L}\left(x,\frac{Q^2}{\mu^2} \right) \otimes 
A^{ij}\left(x,\frac{m_q^2}{\mu^2}\right).
\end{eqnarray}

The leading order (LO) heavy flavor corrections were calculated  for neutral 
and charged current interactions in \cite{LO} and numerically at next-to-leading order (NLO) in 
\cite{NLO}. Analytic results 
at NLO have been obtained in complete form for the non--singlet and pure singlet corrections in 
\cite{Buza:1995ie,Buza:1996xr,Blumlein:2016xcy,Blumlein:2019qze} and in the asymptotic case $Q^2 \gg m_q^2$ 
in Refs.~\cite{Buza:1995ie,Buza:1996xr,Buza:1997mg,Buza:1996wv,NLO2,Blumlein:2014fqa}.

At 3--loop order at present only results in the asymptotic case have been calculated. In the single mass 
case the unpolarized OMEs for all OMEs $\propto N_F$ and the complete results for $A_{qq,Q}^{\rm NS}$, 
$A_{qg,Q}$, 
$A_{qq,Q}^{\rm PS}$, $A_{Qq}^{\rm PS}$, $A_{gq,Q}$\footnote{In the non--singlet case we have also 
calculated the OMEs for transversity.} have been calculated in Refs.~\cite{Ablinger:2010ty,
Blumlein:2012vq,Ablinger:2014vwa,Ablinger:2014nga,Ablinger:2014lka} and those $\propto T_F^2$ for $A_{gg,Q}$ in 
\cite{Ablinger:2014uka}. All logarithmic corrections were computed in \cite{Behring:2014eya}.
For the pure $N_F$ contributions and both in the non--singlet case and for $A_{gq,Q}$ the OMEs can be 
expressed by harmonic sums \cite{HSUM} or harmonic polylogarithms \cite{Remiddi:1999ew} only. In the pure 
singlet case also generalized harmonic sums \cite{Moch:2001zr,Ablinger:2013cf} contribute. In $z$ space 
harmonic polylogarithms of argument $z$ do not span the result, unless one allows also for the 
argument 
$1-2 z$. For the $T_F^2$ terms  of $A_{gg,Q}$ also nested finite binomial sums contribute, leading to 
root--valued iterated integrals \cite{Ablinger:2014bra}. Rather involved binomial structures also occur in 
the case of $A_{Qg}$ \cite{Ablinger:2015tua}. The OME $A_{Qg}$ also contains iterative non--iterative 
integrals \cite{Ablinger:2017bjx}, containing complete elliptic integrals. The first order factorizing 
contributions to $A_{Qg}$ have been calculated in \cite{Ablinger:2017ptf}.
Phenomenological predictions in the non--singlet case have been given for the the structure functions 
$xF_3$ and $F_L^{W^+-W^-}(x,Q^2)$ and $F_2^{W^+-W^-}(x,Q^2)$ in \cite{Behring:2015roa,Behring:2016hpa}.

In the polarized case single mass contributions have been computed for the OMEs $A_{qq,Q}^{\rm NS}$,
$A_{qg,Q}$, $A_{qq,Q}^{\rm PS}$, $A_{Qq}^{\rm PS}$ and $A_{gq,Q}$ in 
Refs.~\cite{Ablinger:2014vwa,Ablinger:2019etw,Blumlein:2021xlc,Behring:2021asx} and for all logarithmic 
contributions in \cite{Blumlein:2021xlc}. Here the same mathematical structures as in the unpolarized 
case contribute. Phenomenological predictions in the non--singlet case for $g_1(x,Q^2)$ were given in 
\cite{Behring:2015zaa}.

The two--mass corrections contributing from 3--loop order onward were calculated in the unpolarized case
for $A_{qq,Q}^{\rm NS}, A_{qg,Q}$ in \cite{Ablinger:2017xml}, $A_{gg,Q}$ \cite{Ablinger:2018brx} and 
$A_{Qq}^{\rm PS}$ \cite{Ablinger:2017xml} and analogously in the polarized case \cite{Ablinger:2019gpu,
Ablinger:2019gpu,Ablinger:2020snj,Behring:2021asx}. The analytic results in $z$ space can be expressed by 
iterative integrals over root--valued letters, also parameterized by the mass ratio $m_c^2/m_b^2$.
 
Because of the missing hierarchy between $m_c^2$ and $m_b^2$, one has to decouple both these heavy quark 
effects together in a variable flavor number scheme \cite{Buza:1996wv}, generalizing the 
single--mass 
variable flavor number scheme \cite{Ablinger:2017xml,Blumlein:2018jfm}. 

Further calculations concern the missing terms for the OMEs $A_{gg,Q}$ in the single mass case and 
for $A_{Qg}$ in the single and two--mass case. For $A_{Qg}$ there is a first phenomenological 
representation in \cite{Kawamura:2012cr} based on only five Mellin moments calculated by us in 
Ref.~\cite{Bierenbaum:2009mv}.
\section{Calculation of the 3--loop OMEs}
\label{sec:3}

\vspace*{1mm}\noindent
A survey on the calculation methods for the 3--loop massive OMEs has been given in Ref.~\cite{Blumlein:2019svg}.
Massive OMEs contain, beyond the usual Feynman rules, those for the twist--2 local operators, 
cf.~\cite{Bierenbaum:2009mv}. To apply integration-by-parts (IBP) techniques \cite{vonManteuffel:2012np} one 
needs to resum these operators into propagators \cite{Ablinger:2014yaa}. In the topological simple cases we 
perform the calculation of the 3--loop integrals directly, using hypergeometric techniques \cite{HYPERG}.
More complex integrals are calculated using the method of first order factorizing differential 
and difference equations \cite{Ablinger:2015tua,Ablinger:2018zwz}. From the 
IBP reductions one obtains 
systems of differential equations 
which can be mapped into systems of difference equations, allowing to 
calculate a large number of Mellin moments for the master integrals and the OMEs by using the method of
arbitrary large moments \cite{Blumlein:2017dxp}. Having generated a sufficient number of moments the method 
of guessing \cite{GUESS,Blumlein:2009tj} allows to find the difference equations for the respective color and zeta factors 
\begin{figure}[H]
\includegraphics[width=0.49 \linewidth]{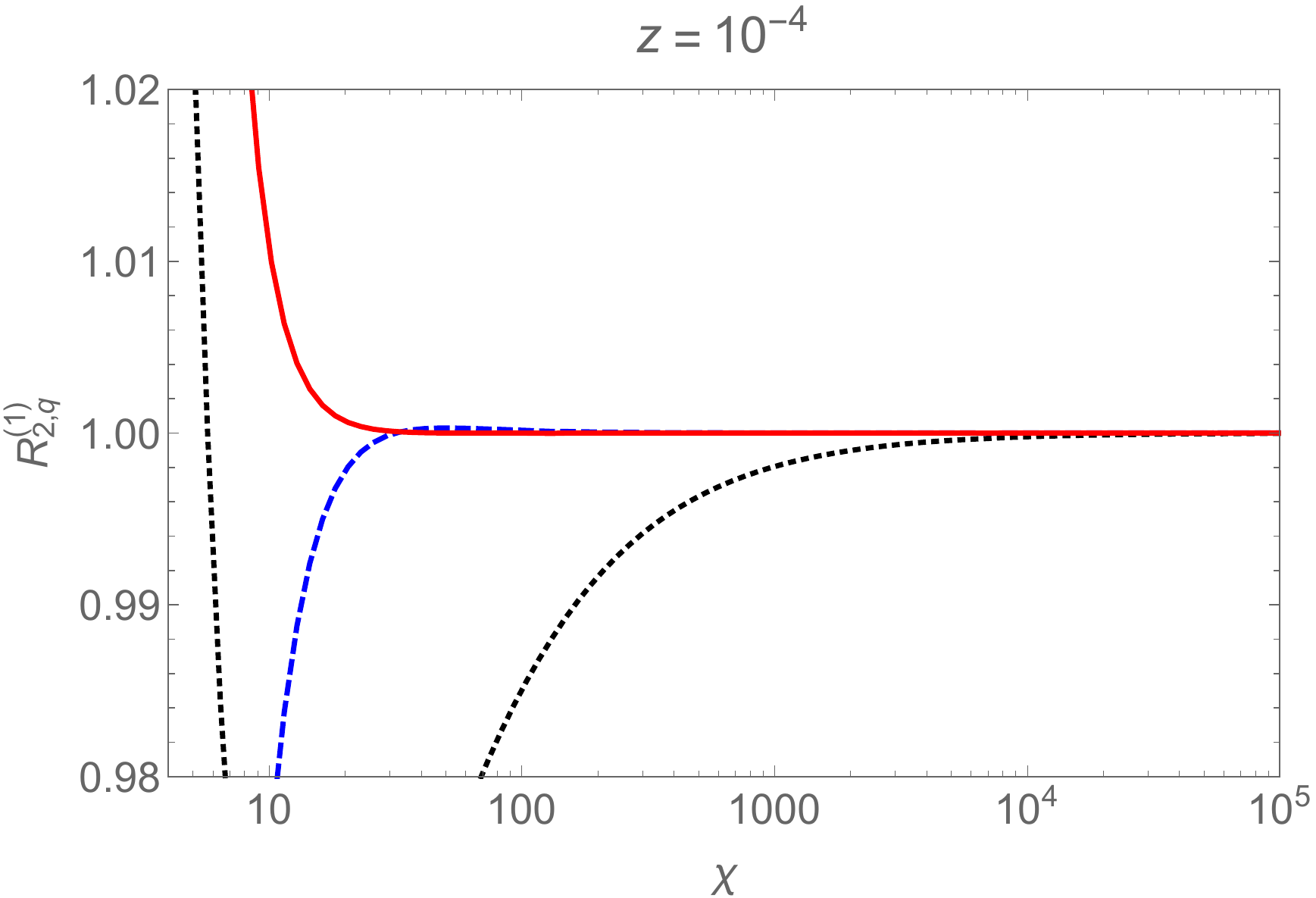}
\includegraphics[width=0.49 \linewidth]{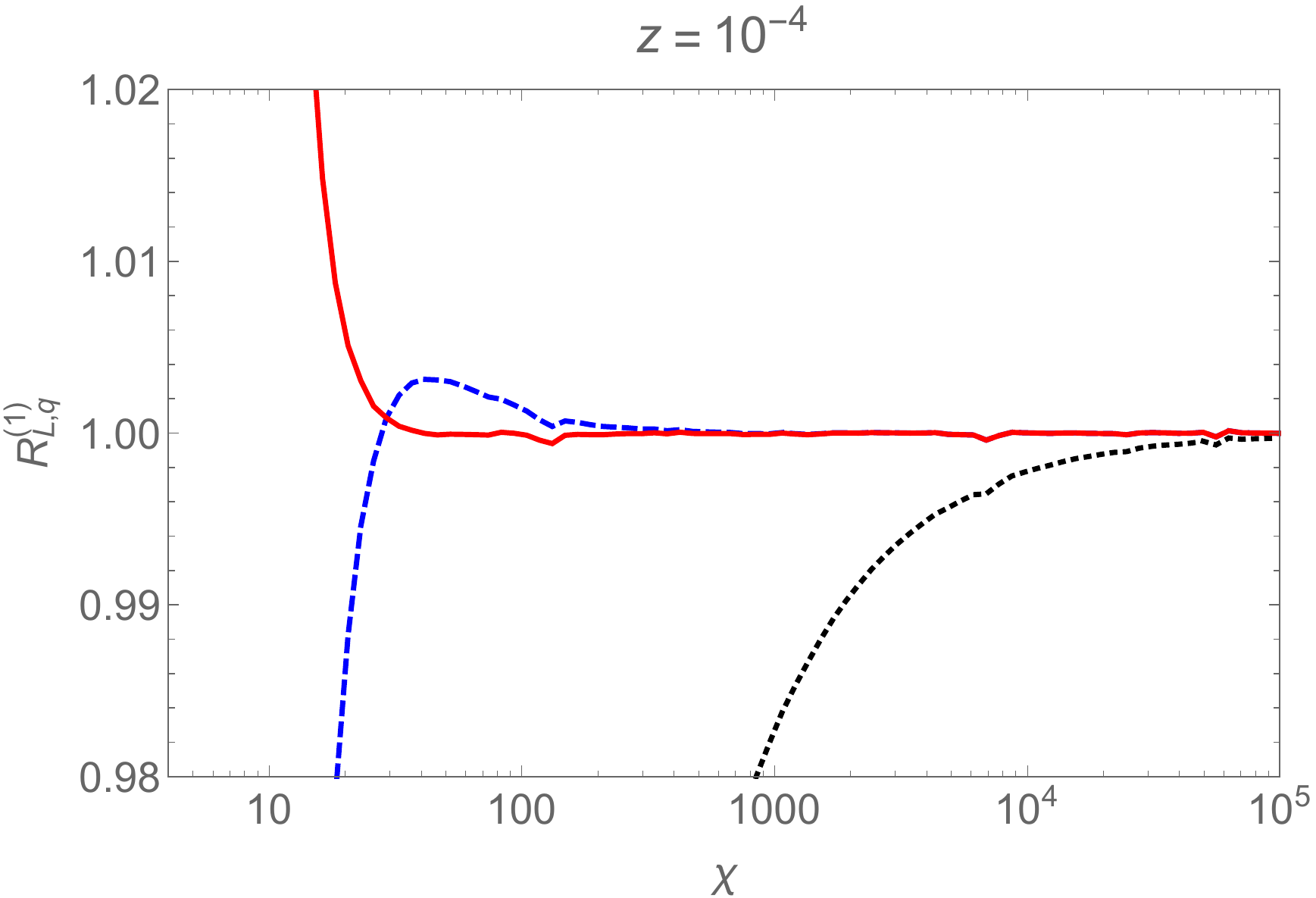}
\includegraphics[width=0.49 \linewidth]{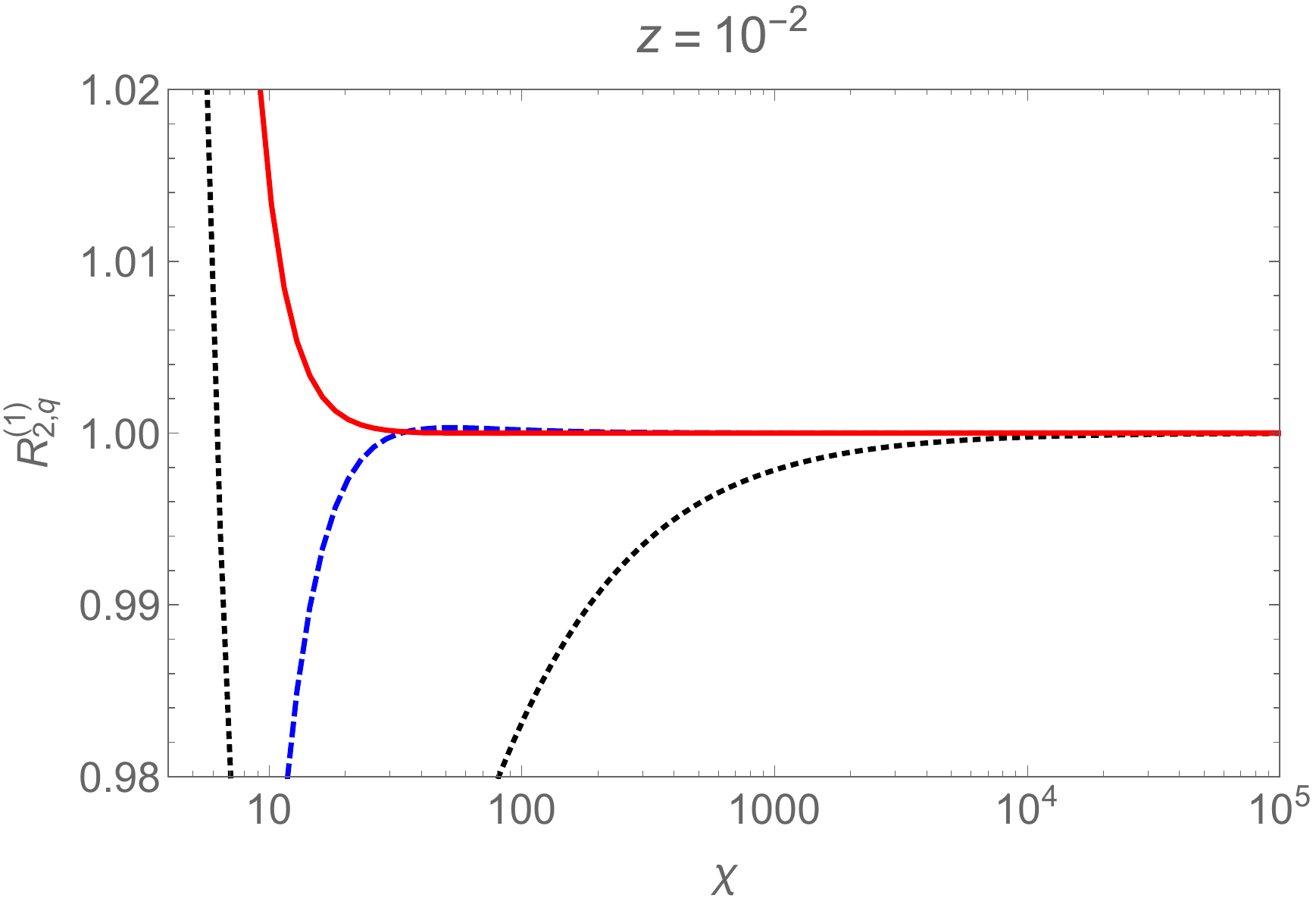}
\includegraphics[width=0.49 \linewidth]{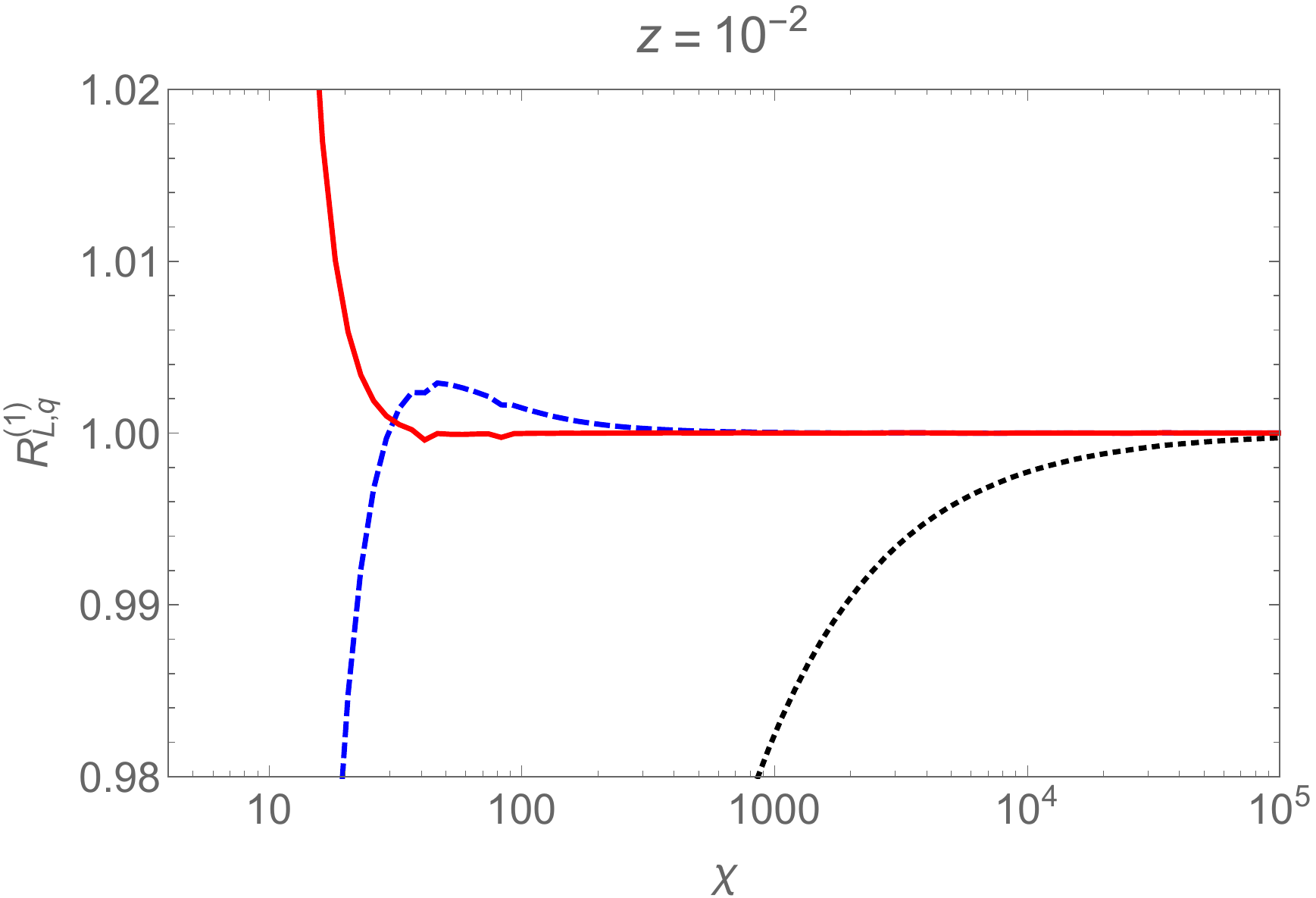}
\includegraphics[width=0.49 \linewidth]{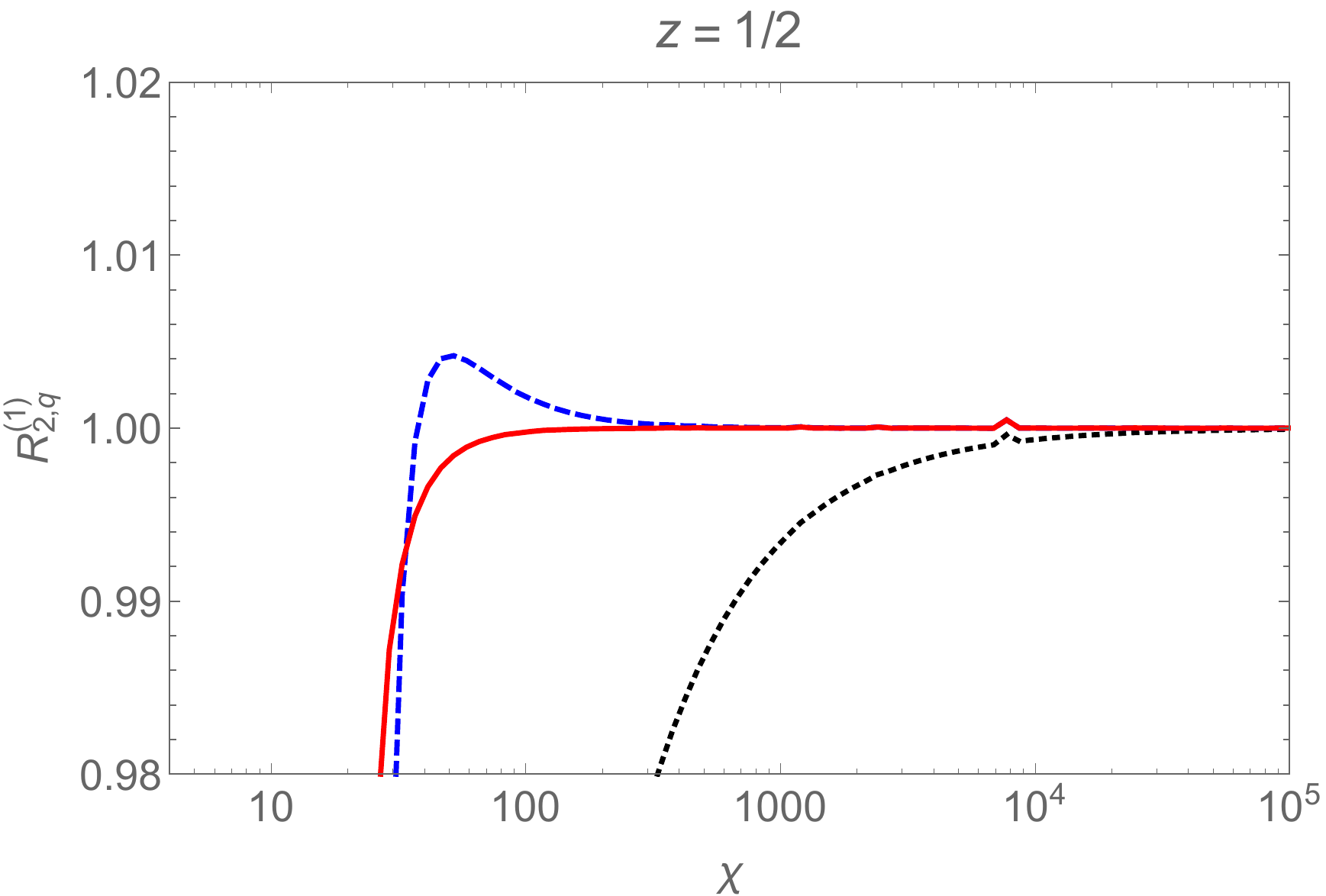}
\includegraphics[width=0.49 \linewidth]{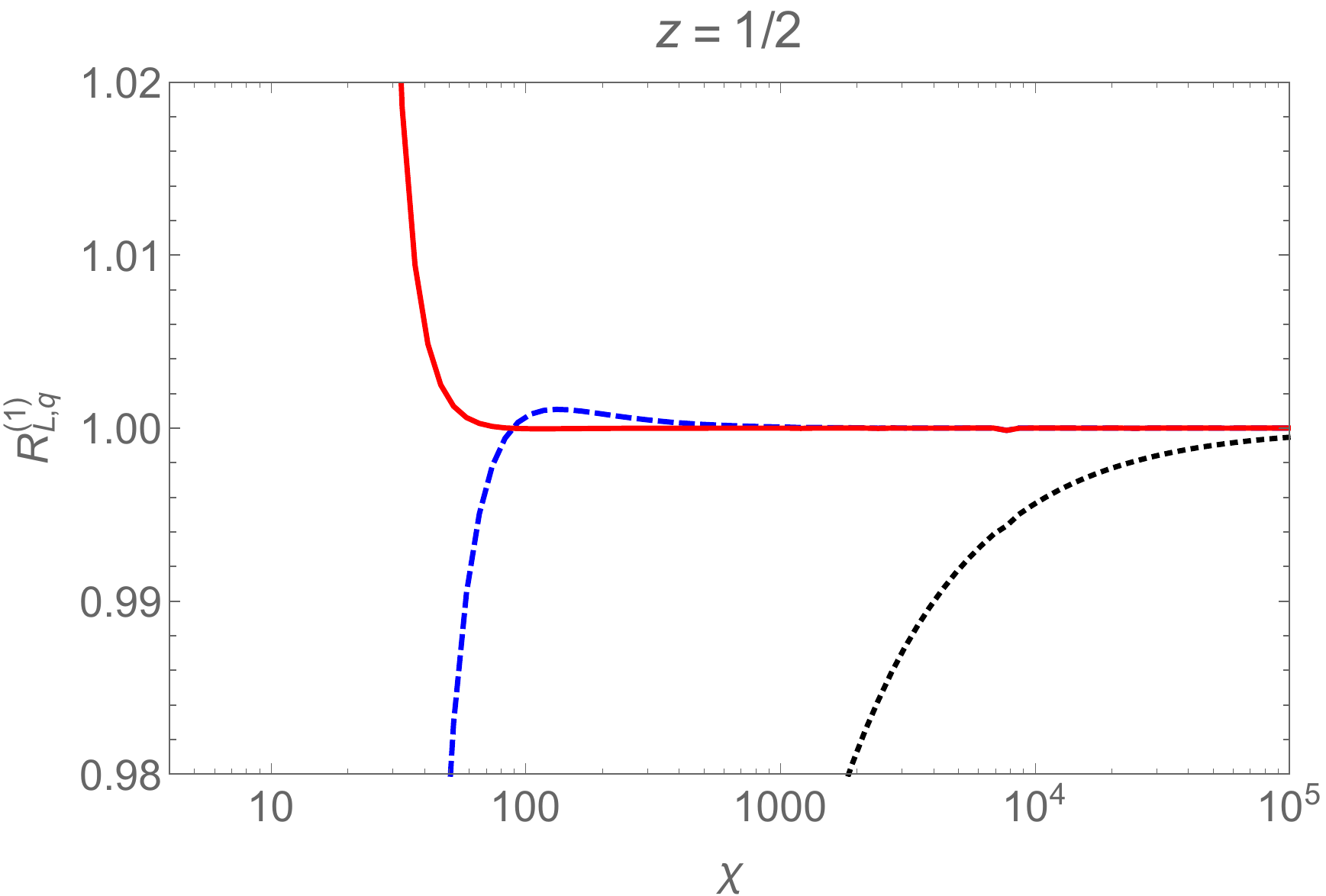}
\caption{\small \sf The ratios of the power expanded structure function to the complete structure function, $R_{2,q}^{(1)}$ 
(left) and $R_{L,q}^{(1)}$ (right), as a function of $\chi = 1/\kappa = Q^2/m^2_q$ for different values of $z$ 
gradually 
improved with 
$\kappa$ suppressed terms. Dotted lines: asymptotic result; dashed lines: $O(m^2_q/Q^2)$ improved; solid lines : 
$O((m^2_q/Q^2)^2)$ improved. From Ref.~\cite{Blumlein:2019qze}.
\label{fig:RLandR2b}}
\end{figure}

\noindent
of the different OMEs. The number of these moments in the case of $A_{Qg}^{(3)}$ is very large. Already for 
the $T_F^2$ terms one needs $\sim 7500$ moments \cite{Ablinger:2017ptf}.

If the difference equations obtained factorize to first order the difference ring techniques 
\cite{DIFFRING} implemented in the package {\tt Sigma} \cite{SIGMA} are sufficient to find the final 
solution in Mellin $N$ space. The $z$ space solutions can be obtained by the techniques implemented in 
the package {\tt HarmonicSums} \cite{HARMSU,HSUM,Remiddi:1999ew,Ablinger:2013cf,Ablinger:2014bra} 
in terms of iterated integrals over certain alphabets, which are found algorithmically. All massive 3--loop OMEs 
except those in the single and two--mass case of $A_{Qg}^{(3)}$ belong to this class and have been solved by now. 

In the following we describe some recent results in calculating single and two--mass contributions at the 
2-- and  3--loop level.
\section{Single mass contributions}
\label{sec:4}
\vspace*{1mm}
\noindent
We consider first the successive inclusion of power corrections in $m_q^2/Q^2$ to the asymptotic result
for the heavy flavor contributions to the structure functions $F_2(x,Q^2)$ and $F_L(a,Q^2)$ at NLO.
The power corrections are of clear importance for $F_L$, since the pure asymptotic terms describe
the heavy flavor corrections for $Q^2/m_q^2 \gsim 800$ only. Also in the case of $F_2$ for larger values of 
$z$ an improved description is obtained. The results for the pure--singlet contribution to the polarized structure 
function $g_1(x,Q^2)$ \cite{Blumlein:2019zux} are similar. The inclusion of the power corrections in the pure--singlet 
case in the unexpanded expressions leads to new iterative integrals over a corresponding alphabet, 
cf.~\cite{Blumlein:2019qze,Blumlein:2019zux}. Expanding in the ratio 
$m_q^2/Q^2$ leads to harmonic polylogarithms again. The situation is simpler in the non--singlet case, where classical 
polylogarithms with more complicated arguments suffice in the general case, cf. e.g.~\cite{Blumlein:2016xcy}. Note that
the corresponding Wilson coefficients are not the ones given in \cite{Buza:1995ie}, but need an extension.

The most important calculation for the future consist in the analytic computation of the OME $A_{Qg}^{(3)}$.
The terms $\propto N_F$ were calculated in \cite{Ablinger:2010ty}. Based on 1000 Mellin moments all contributions, but the 
pure rational and $\zeta_3$ terms were calculated, since these can be obtained from difference equations which are factorizing 
at first order, which is not the case for the former terms, \cite{Ablinger:2017ptf}. Based on 8000 moments we obtained the 
difference equations for all missing terms $\propto T_F^2$. In the $T_F$-case one will need more moments to find the 
corresponding difference equations. In the $T_F^2$ case the difference equations have the following 
characteristics for 
degree {\sf d} and order {\sf o}:
\begin{eqnarray*}
T_F^2 C_A  &:&  {\sf (d;o)} = (1407; 46)
\\
T_F^2 C_A  \zeta_3 &:&  {\sf (d;o)} = (323; 24)
\\
T_F^2 C_F  &:&  {\sf (d;o)} = (654; 27)
\\
T_F^2 C_F \zeta_3 &:&  {\sf (d;o)} = (283; 14).
\end{eqnarray*}
The first difference equation is  more voluminous than the largest occurring in guessing the largest contribution to the 
3--loop massless Wilson coefficient in Ref.~\cite{Blumlein:2009tj}. For the 3--loop massive form factor, 
\cite{Blumlein:2019oas}, one difference equation of ${ \sf (d,o)} = (1324; 55)$ has been obtained.

The separate analysis of the difference equations for the rational and $\zeta_3$ $T_F^2$ cases showed, that the associated 
differential equations develop exponential singularities in the region $z \in$ $[0,1]$, although the complete solution
is regular. Indeed, the differential equation for the purely rational term develops a $\zeta_3$ factor 
asymptotically such, that both the 
singularities cancel, which requires to deal with both difference equations at once, despite the fact, that one can
establish them only separately. 

Starting from the difference equations in Mellin $N$ space one may find Laurent series solutions of the associated
differential equations around $z_0 = 1$ up to a finite upper power in $z$. This expansion is possible also around any 
other regular point $z_0 \in [0,1]$, \cite{ABS21}. All these expansions have a finite convergence radius and several 
expansions are necessary to map out the interval $z \in [0,1]$ in terms of overlapping expansions. One obtains an 
approximate analytic solution, containing high--precision numerical constants in part, in this way, which may be 
tuned to any 
accuracy. Given the fact that also all the 
other special functions need numerical representations, the numerical representation in this case is already obtained.
The generation of a high number of moments for the $T_F$ terms is underway.

In Figure~2 we summarize the different contributions to the currently known charm quark QCD corrections up to 
3--loop order 
to the structure function $F_2$ at the scale $Q^2 = 100~\GeV^2$.

\begin{figure}[H]\centering
\includegraphics[width=0.60 \linewidth]{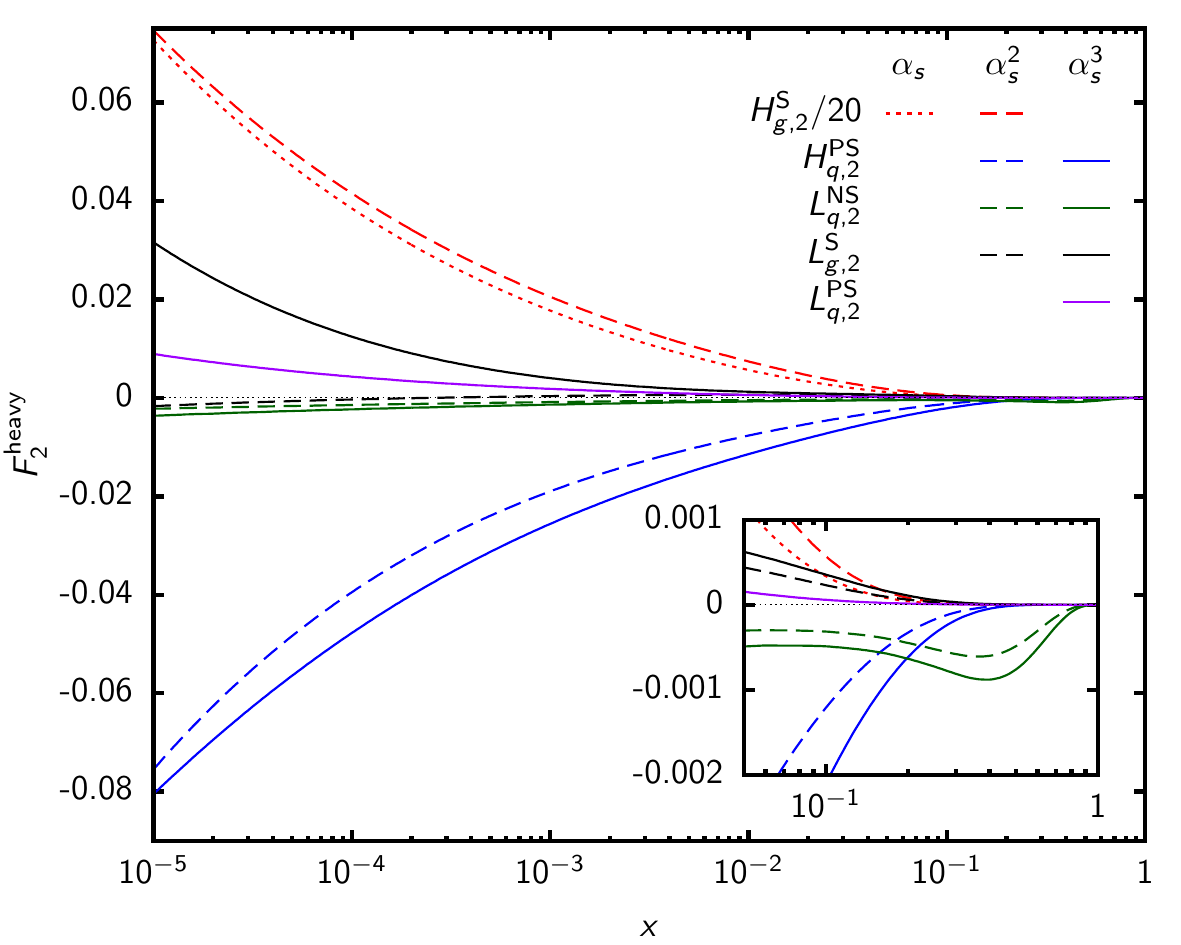}
\caption{\small \sf 
The different single mass heavy flavor contributions to the structure function $F_2$
at $Q^2 = 100 \GeV^2$.
From \cite{Ablinger:2016kgz}.}
\end{figure}

\section{Two-mass contributions}
\label{sec:5}

\vspace*{1mm}\noindent
Irreducible two--mass contributions emerge first at 3--loop order and lead to a change of the variable flavor 
number scheme
\cite{Ablinger:2017xml}. Reducible contributions imply two--mass contributions already at NLO 
\cite{Blumlein:2018jfm}. In 
Figure~3 we illustrate the 2--mass effects on the singlet and bottom quark contribution. In the singlet case the 
effect 
amounts to 1\% and in the case of the $b$-quark distribution of 4--5\%.
\begin{figure}[H]
\includegraphics[width=0.49 \linewidth]{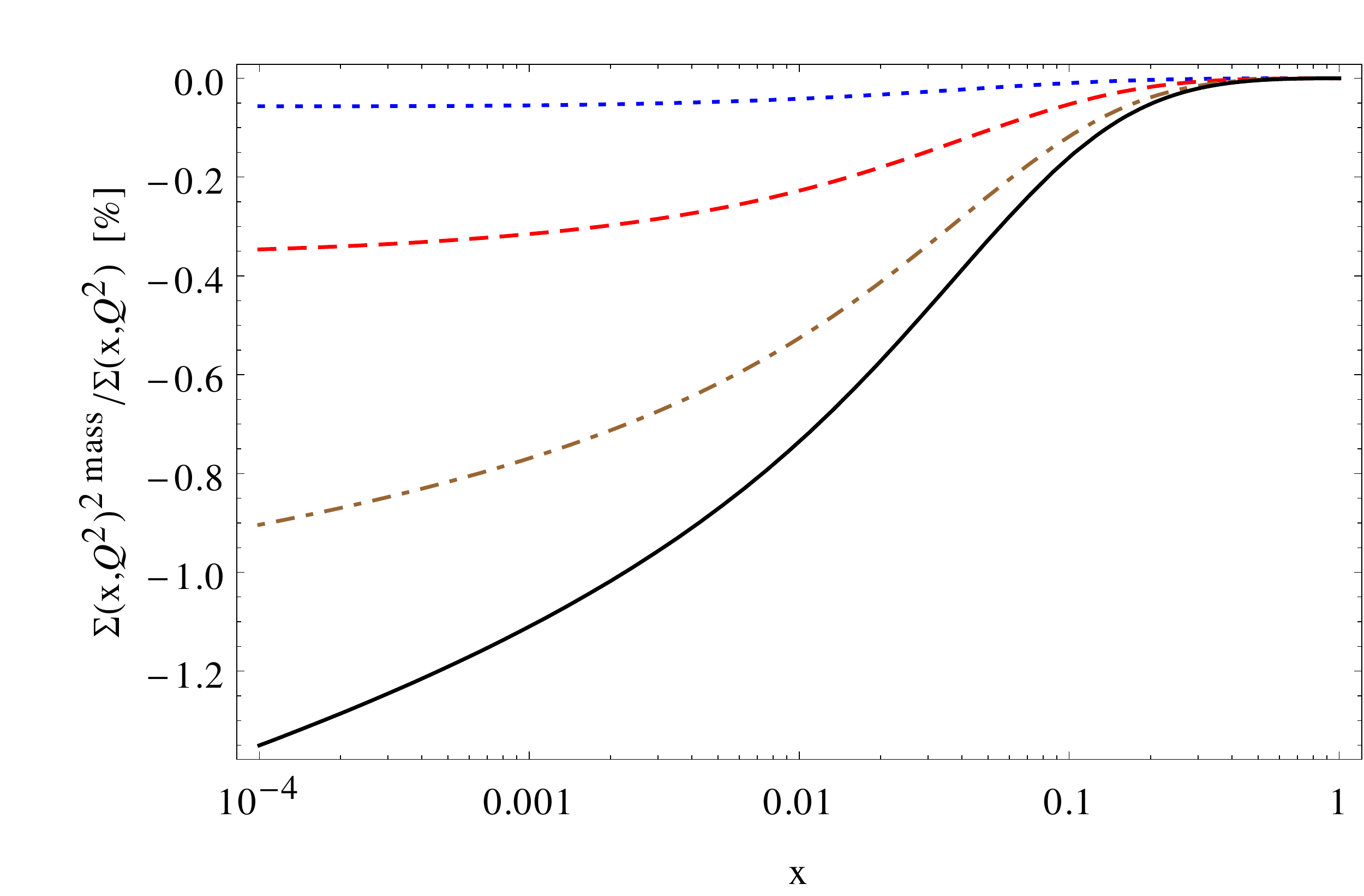}
\includegraphics[width=0.49 \linewidth]{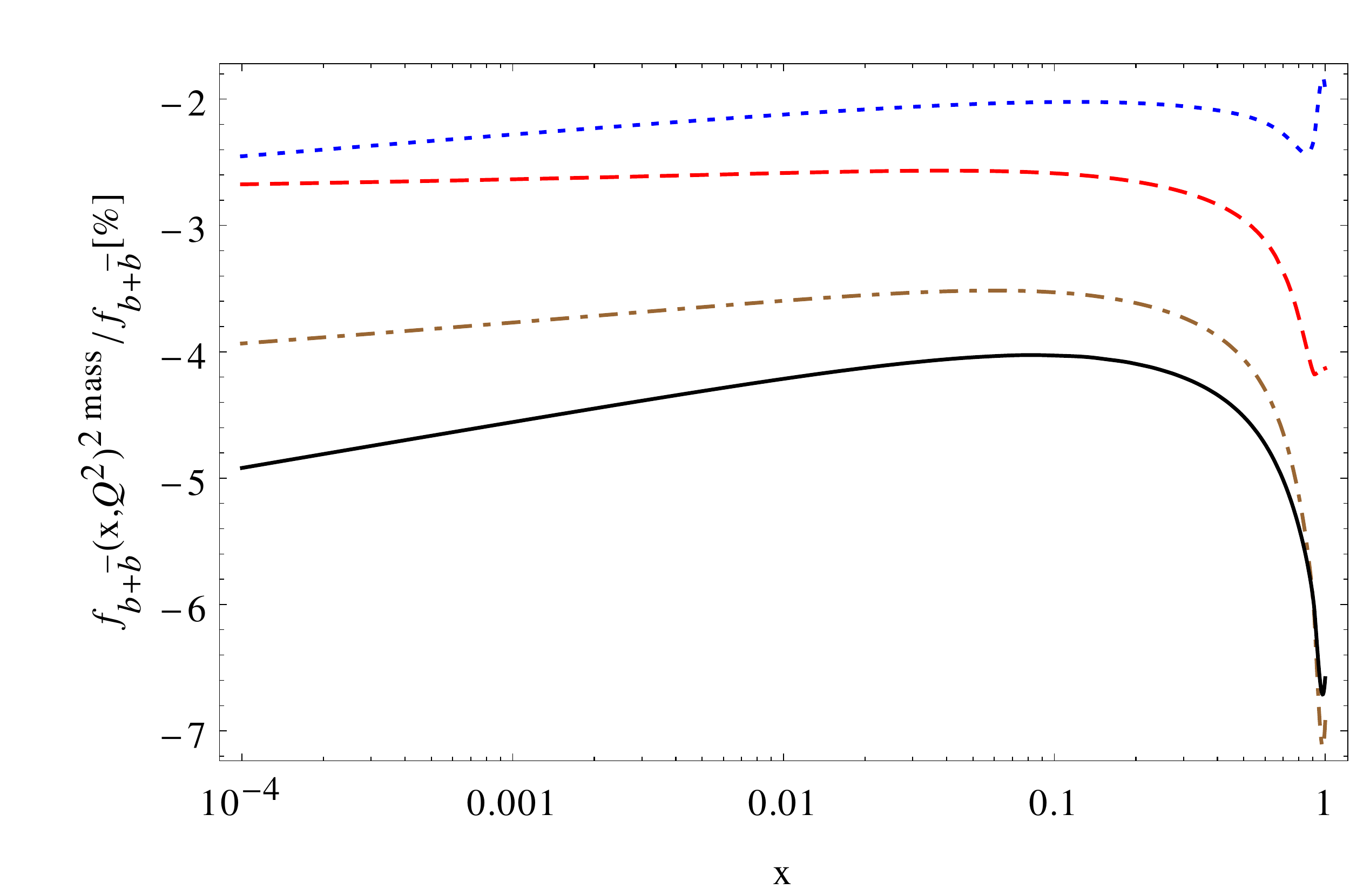}
\caption{\small \sf 
The ratio of the 2-mass contributions to the singlet parton distribution $\Sigma(x,Q$
        in the on-shell scheme.
        Dash-dotted line:  $Q^2 =   30~\text{GeV}^2$;
        {Dotted line:  $Q^2 =   30~\text{GeV}^2$;}
        {Dashed line:  $Q^2 =   100~\text{GeV}^2$;}
        {Dash-dotted line:  $Q^2 =   1000~\text{GeV}^2$;}
        Full line:  $Q^2 =   10000~\text{GeV}^2$. For the PDFs the \texttt{NNLO} 
variant of ABMP16 with $N_f = 3$ flavors was used \cite{Alekhin:2017kpj}. Form \cite{Blumlein:2018jfm}.}
\end{figure}

We also have calculated the next-to-next-to leading order (NNLO) heavy flavor corrections to the flavor non--singlet 
structure functions $F_2^{\rm NS}(x,Q^2)$ and $g_1^{\rm NS}(x,Q^2)$ in the case of scheme--invariant evolution
\cite{Blumlein:2021lmf} shown in Figure~4. Here one considers the evolution 
\begin{eqnarray}
F_2^{\rm NS}(x,Q^2) = E_{\rm NS}(x,Q^2,Q_0^2) \otimes F_2^{\rm NS}(x,Q^2_0)
\end{eqnarray}
from a starting scale $Q_0^2$ to $Q^2$, where $E_{\rm NS}(x,Q^2,Q_0^2)$ denotes a scheme--invariant evolution operator
and the input distribution function $F_2^{\rm NS}(x,Q^2_0)$ is measured experimentally. The evolution operator also depends
on the masses $m_c$ and $m_b$.
\begin{figure}[H]
        \centering
        \includegraphics[width=0.49\textwidth]{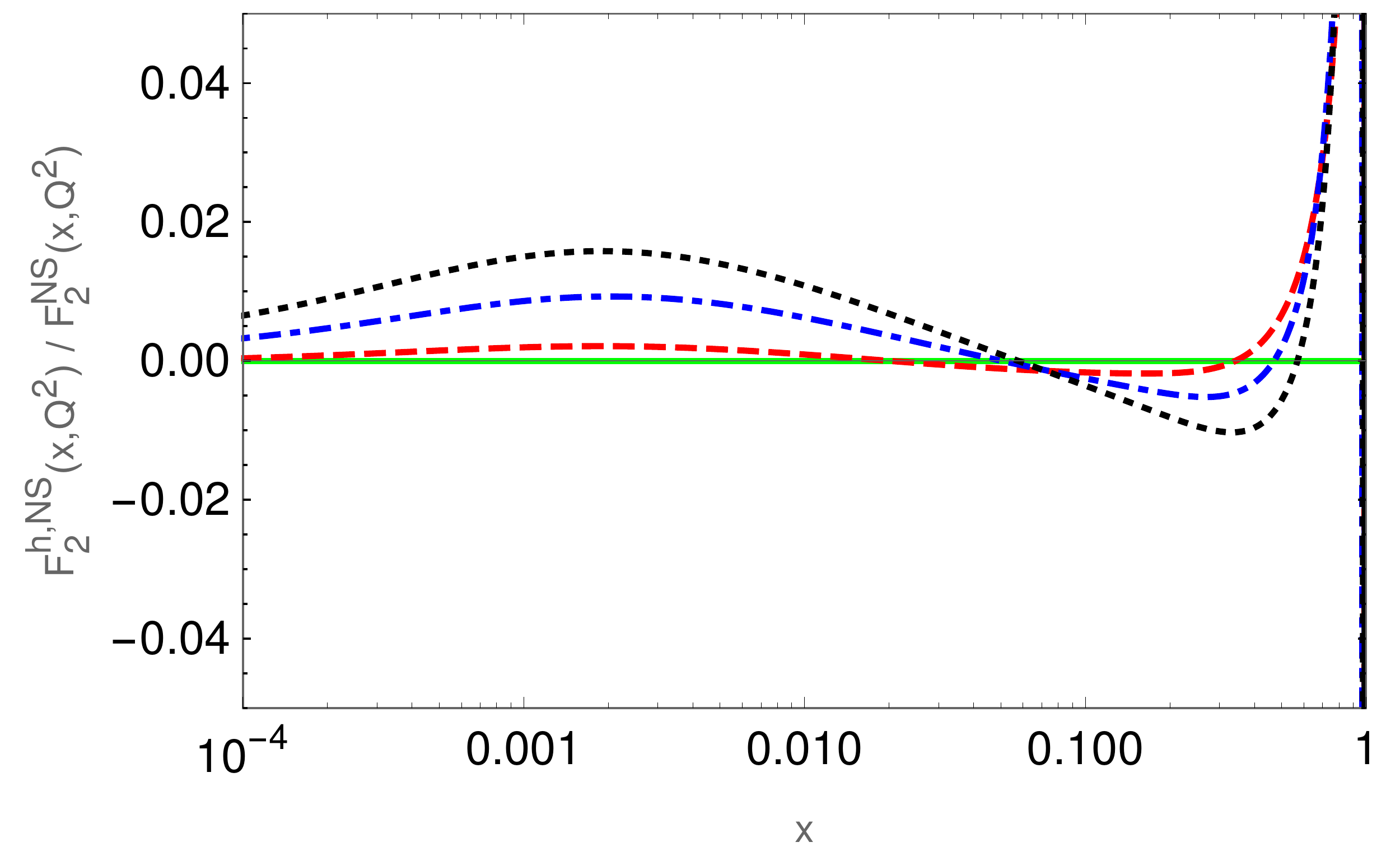}
        \includegraphics[width=0.49\textwidth]{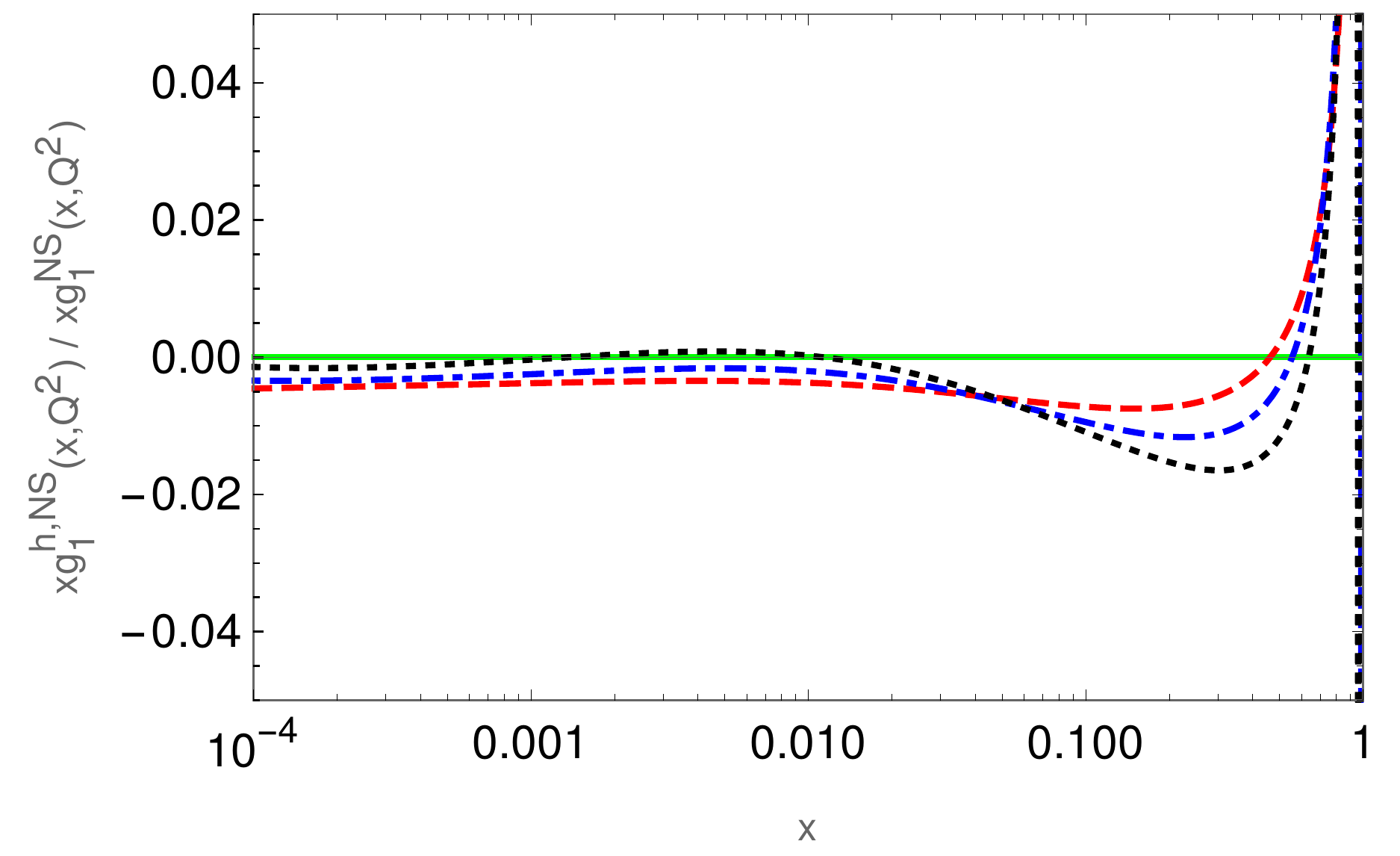}
        \caption{\sf Left:~The relative contribution of the heavy flavor contributions due
to $c$ and $b$ quarks to the structure function $F_2^{\rm NS}$ at N$^3$LO;
dashed lines: $100~\GeV^2$;
dashed-dotted lines: $1000~\GeV^2$;
dotted lines: $10000~\GeV^2$. Right:~The same
for the structure function $xg_1^{\rm NS}$ at N$^3$LO. From \cite{Blumlein:2021lmf}.
\label{fig5}}
\end{figure}

\noindent
These corrections amount to $O(1\%)$ and are important in future factorization--scheme invariant measurements 
from the scaling violations of these structure functions at high luminosity facilities like the EIC or the LHeC.
\section{Conclusions}
\label{sec:6}

\vspace*{1mm}\noindent
Most of the massive 3–loop OMEs have been calculated in the single and two--mass case. They contribute to the 
(two--mass) variable flavor number scheme and the heavy flavor Wilson coefficients in unpolarized and 
polarized deep--inelastic scattering in the asymptotic region $Q^2 \gg m_q^2$. All quantities which can be described by 
first order factorizing difference equations have been computed. For the remaining OMEs the determination of their 
difference equations is underway. A method has recently been developed that can solve these equations as well.
In all the computations extensive use has been made of the methods of arbitrary high Mellin moments, the method of 
guessing, and of difference ring theory to solve the respective physical problems. In course of these computations
a series of different function spaces has been found to perform different intermediary steps of the calculation 
and to find a minimal analytic representation of the final results.
 
At two--loop order analytic results have been derived for the non--singlet and pure--singlet Wilson coefficients in the 
whole kinematic region. The 2--mass corrections are quantitatively as important as the $O(T_F^2)$ contributions in 
the single 
mass case. The $O(T_F)$ contributions to the 3-loop anomalous dimensions have been calculated as by-product of the 
calculation of the massive OMEs and they agree with the results of previous calculations. The calculation methods 
have also being applied
to higher order massive QED corrections for the initial state radiation to observables in $e^+e^-$ annihilation~\cite{EPEM}.

\vspace*{4mm} \noindent
{\bf Acknowledgment.}~
This project has received funding from the European Union's Horizon 2020 research and innovation programme
under the Marie Sk\l{}odowska--Curie grant agreement No. 764850, SAGEX and from the Austrian Science Fund (FWF)
grant SFB F50 (F5009-N15). 


\end{document}